# Grid Integration of Gigawatt-Scale AI Data Centers under Connect-and-Manage

Xin Lu, *Senior Member*, *IEEE*, Qianwen Xu, *Senior Member*, *IEEE*

***Abstract*— Emerging connect-and-manage interconnection practices allow gigawatt-scale artificial intelligence data centers (AIDCs) to connect to the transmission network without prior network upgrades, at the cost of real-time curtailment during grid stress. This paper formalizes the resulting AIDC–transmission system operator (TSO) coordination as a sequential request–acceptance protocol with an explicit curtailment variable and a strict information boundary between the two parties. Physical models are developed on both sides of the point of common coupling: the AIDC is decomposed into frontier training, batch training, and inference serving subclasses sharing on-site battery energy storage, capturing differentiated temporal flexibility; the transmission network is modeled via DC power flow with generator constraints and budget-constrained demand uncertainty. Because the TSO's acceptance mapping is opaque to the AIDC, a three-layer hierarchical architecture is formulated in which a learning-based planning layer generates power requests, the TSO evaluates each request through a robust acceptance mechanism, and a single-step execution optimizer enforces internal feasibility under the realized power budget. Case studies with a gigawatt-scale AIDC on the IEEE 39-bus system with Australian market data show that the framework reduces curtailment from 9.1% to 2.8% while preserving 98.1% frontier training workload, that batch training acts as the primary grid-elastic resource with the largest throughput swing during peak demand, and that the on-site battery provides curtailment buffering through active discharge and charge deferral.***

## I. Introduction

The explosive growth of generative artificial intelligence (AI) has turned AI data centers (AIDCs) into one of the fastest-growing categories of grid-connected loads in modern power systems. A single training facility can draw several hundred megawatts to well above one gigawatt of electricity, and leading industry reference designs already target near-gigawatt AI factories [1]. Global data center capacity demand is projected to nearly triple from approximately 82 GW in 2025 to 219 GW by 2030, with AI workloads accounting for roughly 70% of the total [2]. The International Energy Agency projects that worldwide electricity consumption from data centers will more than double to 945 TWh per year by 2030 [3]. Such growth is fundamentally misaligned with the multi-year timescale of transmission reinforcement, creating a structural gap between the speed of AIDC deployment and the speed of conventional grid expansion [4].

On 16 January 2026, the PJM Interconnection Board announced a connect-and-manage framework for new large loads, under which incremental demand that does not bring its own generation may be subject to curtailment prior to the deployment of pre-emergency demand response [5]. Industry demonstrations confirm that AIDC can already modulate power consumption in response to grid signals through software-defined workload orchestration: a field test on a 256-GPU cluster in Phoenix reduced power usage by 25% for three hours during peak demand while maintaining AI quality-of-service guarantees [6], and the 96 MW Aurora AI Factory jointly developed by NVIDIA, EPRI, Digital Realty, and PJM demonstrated real-time response to time-varying operating limits from the grid operator [7].

Connect-and-manage is thus no longer a prospective policy concept; it is an emerging operational regime that transforms the AIDC–grid relationship from a one-shot interconnection decision into a continuous sequence of real-time interactions between the load and the system operator. The interaction itself has no established representation in the literature, and its introduction reshapes the AIDC's operational decision-making in ways that existing models do not capture.

Existing studies on data center–grid coordination adopt one of two interaction paradigms. In the first, the grid operator issues time-varying power limits or price signals to the data center, which responds as a passive demand-response resource [8, 9]. In the second, data centers participate in energy or ancillary-service markets through bid-based mechanisms [10, 11]. In both cases, the power exchange is set through a single mechanism, namely a unilateral signal in the first paradigm and a market-clearing rule in the second, rather than through repeated bilateral interaction at each operating interval. Neither paradigm captures a setting in which the load and the operator make interdependent decisions and the realized power exchange depends on both parties' actions. The protocol-level structure of such an interaction, including the timing, the interface variables, and the information available to each party, therefore becomes a modeling question in its own right that has received little attention.

The physical modeling of data centers in the power systems literature has focused primarily on traditional IT facilities with relatively homogeneous and freely adjustable server loads [12]. Several studies incorporate thermal dynamics [13] or workload migration across geographically distributed sites [14], but treat the computing load as a single controllable block. This simplification is increasingly at odds with modern AIDCs, where different workload categories coexist within a single facility and exhibit markedly different tolerance to interruption, deferral, and throughput reduction [15]. On-site battery energy storage, increasingly deployed in hyperscale facilities as a buffer against supply variability [16], has not been integrated into AIDC-level operational models. On the grid side, studies that co-optimize data center operation with transmission dispatch typically grant the data center full visibility into network topology, line ratings, and generator costs [17], an assumption that breaks down whenever the grid-side feasibility assessment is an internal process whose outcome the data center can observe but whose mechanism it cannot replicate.

The decision-making problem that couples a large flexible load with a transmission operator's security-constrained dispatch resembles, at first glance, a bilevel program. Bilevel optimization has been widely applied to generator-grid and storage-market interactions [18], but these applications assume the upper-level agent can model or approximate the lower-level problem. When the lower-level agent's objective, constraints, and input data are not disclosed, the bilevel structure becomes inaccessible. Scenario-based stochastic programming faces a similar barrier, as generating plausible scenarios of the operator's response presupposes knowledge of the decision mechanism being modelled [19]. Recent applications of reinforcement learning to energy system scheduling [20] demonstrate that learning-based methods can handle partial observability, but these studies address settings with substantially simpler load models and do not confront the combination of heterogeneous internal flexibility, grid-side robustness requirements, and protocol-mediated information constraints that large-load interconnection introduces.

Together, the preceding observations reveal a layered modeling gap. The bilateral AIDC–TSO interaction itself has no formal protocol representation, while existing physical models on both sides of the PCC fail to capture the AIDC's heterogeneous internal flexibility, its on-site energy storage, or the grid-side feasibility assessment that operates as a black box to the AIDC. Even where models are available, the resulting information asymmetry renders standard bilevel and stochastic optimization approaches inapplicable, leaving no established methodology for AIDC decision-making under connect-and-manage. The contributions are threefold:

1. *Connect-and-manage interaction protocol and information boundary.* A sequential request–acceptance protocol formalizes the real-time AIDC–TSO coordination under connect-and-manage, in which the AIDC submits a power request, the TSO returns an accepted exchange with an explicit curtailment variable, and the AIDC executes internal allocation under the realized power budget. The protocol defines the institutional roles of the two parties: request right for the AIDC, curtailment right for the TSO, and establishes a strict information boundary that delineates what each party observes and what remains private.

2. *Physical modeling on both sides of the PCC.* Unlike existing formulations that treat data center computing load as a homogeneous adjustable block, the AIDC is decomposed into three workload subclasses with distinct constraints and asymmetric under-delivery penalties: batch training can be freely interrupted, frontier training requires continuity protection, and inference must track exogenous demand. An on-site BESS is integrated as a power buffer between the grid-accepted exchange and the internal IT load. On the grid side, the transmission network is modeled via DC power flow with a budget-constrained demand uncertainty set to characterize the feasibility region for the TSO's acceptance decision.

3. *Hierarchical decision architecture under information asymmetry.* A three-layer hierarchical architecture organizes the real-time decision sequence. The planning layer maintains a capacity-aware policy that coordinates workload pacing across clusters and anticipates curtailment patterns over the scheduling horizon, producing a power request at each time step. The TSO evaluates the request through a robust acceptance mechanism and returns the accepted exchange. The execution optimizer then solves a single-step optimization that enforces exact power balance, determines BESS dispatch, and reallocates the power budget across clusters when curtailment occurs, tracking the planning intent when the full request is granted and yielding to physical feasibility when it is not.

## II. Connect-and-Manage Interaction Protocol

This section formalizes the operational interaction between the AIDC and the TSO under a connect-and-manage agreement. The protocol defines the sequential timing of decisions, the interface variables exchanged at the point of common coupling, and the information boundary that separates the two parties. The physical models underlying each party's decisions, including the AIDC's internal feasibility region and the TSO's network feasibility region.

### A. Protocol Structure

Under connect-and-manage, the AIDC-TSO interaction proceeds as a sequential protocol at each discrete time step $t \in \{1,2,\dots,T\}$ with resolution $\Delta t$. The protocol consists of three phases per step:

***Request***. The AIDC computes and submits a power request $P_{\text{req}}(t) \geq 0$ to the TSO, reflecting its aggregate operational power needs inclusive of IT load, cooling, and energy storage.

***Acceptance***. The TSO evaluates the request against current network conditions, generator dispatch feasibility, transmission line thermal limits, and demand uncertainty, and returns an accepted exchange:

$$P_{\text{acc}}(t) = P_{\text{req}}(t) - \kappa(t), 0 \leq \kappa(t) \leq P_{\text{req}}(t) \quad (1)$$

where $\kappa(t)$ is the curtailment variable. When network constraints are slack, the TSO accepts the full request ($\kappa(t) = 0$). When constraints bind, the TSO curtails the minimum amount necessary to maintain grid security.

***Execution***. The AIDC receives $P_{\text{acc}}(t)$ and allocates the accepted power across its internal assets, heterogeneous GPU clusters and on-site energy storage, subject to internal physical constraints.

This request–acceptance–execution sequence repeats at each time step throughout the scheduling horizon. The protocol assigns distinct institutional roles: the AIDC holds a request right to submit any non-negative power request up to its nominal connection capacity; the TSO holds a curtailment right to reduce the accepted exchange to any level necessary for network security. The request is treated as a truthful reflection of the AIDC's operational power needs.

### B. Information Boundary

The protocol establishes a strict information boundary between the two parties (illustrated in Fig. 1). The AIDC observes the realized accepted exchange $P_{\text{acc}}(t)$, the curtailment $\kappa(t)$, public system-level signals (electricity spot price $\pi(t)$, aggregate system demand $D(t)$, inference demand $d_{\text{inf}}(t)$), and its own internal state (cluster throughputs, battery state of charge, workload progress). It does not observe network topology, line thermal capacities, the demand uncertainty set, the baseline generator dispatch, or the TSO's objective function. Conversely, the TSO observes only the submitted request $P_{\text{req}}(t)$ and the AIDC's nominal connection capacity; it does not model the AIDC's internal cluster structure, workload state, or energy storage.

This information asymmetry renders the TSO's acceptance mapping: the function from $P_{\text{req}}(t)$ to $P_{\text{acc}}(t)$, a black box to the AIDC, precluding both bilevel optimization and scenario-based stochastic programming. The AIDC must instead learn to anticipate curtailment patterns through repeated interaction. This structural constraint motivates the hierarchical decision architecture developed in Section V.

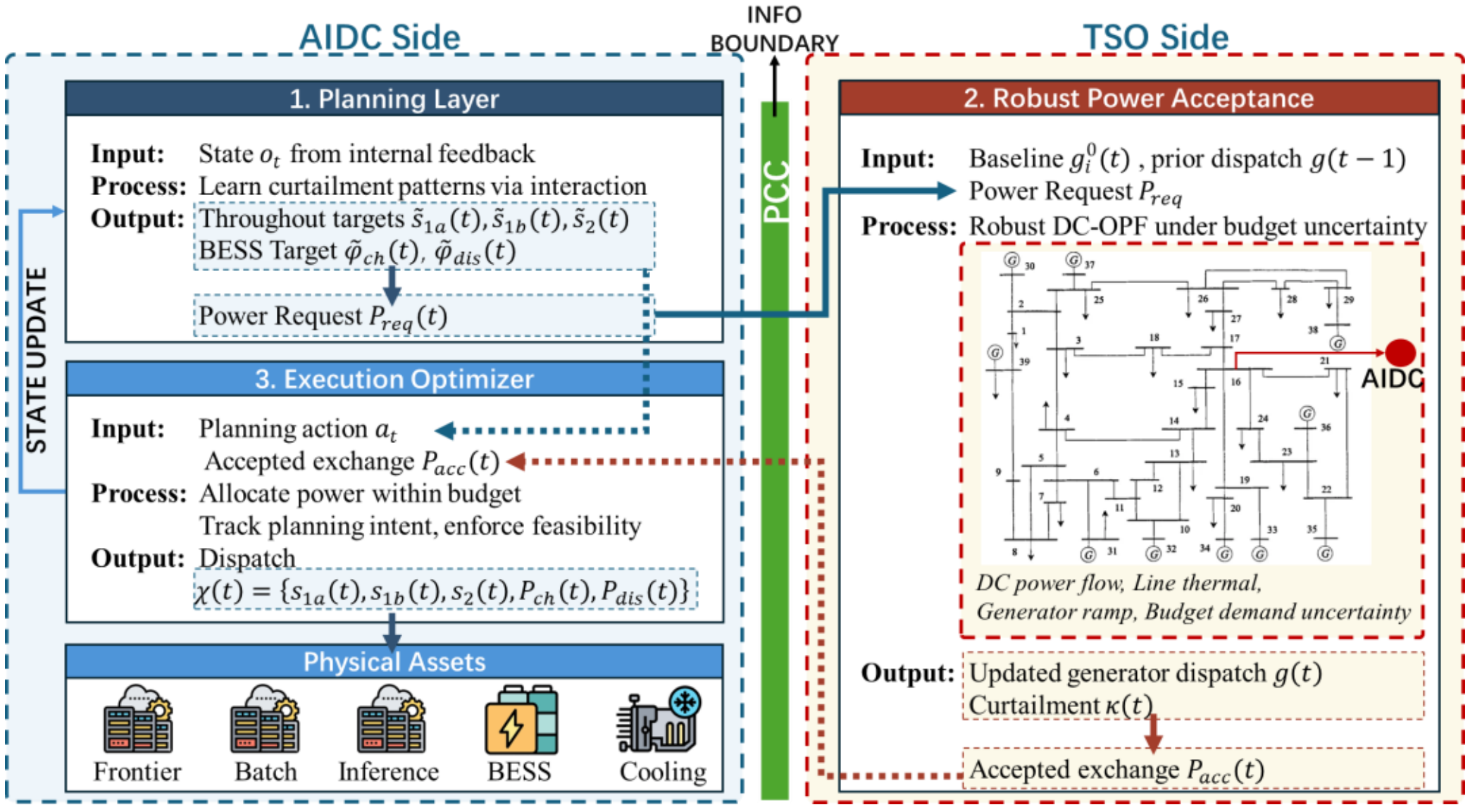

Fig. 1 AIDC–TSO Interaction Framework under Connect-and-Manage

## III. Heterogeneous AIDC Model

A gigawatt-scale AIDC is modeled in this paper as a grid-connected facility composed of two heterogeneous GPU clusters, an on-site battery energy storage system (BESS), and a common cooling subsystem, all connected to the transmission network at the point of common coupling (PCC). The internal electrical architecture has two distinct power paths: the GPU clusters and the BESS are connected on the DC bus behind an internal power conditioning system (IPCS), while the cooling subsystem draws power directly from the AC bus upstream of the IPCS. The two GPU clusters correspond to the two workload categories that dominate hyperscale AI operations in industry practice: Training (Cluster 1) and Inference (Cluster 2), consistent with workload classifications adopted by major hyperscalers [8]. To capture the substantial heterogeneity within training workloads, Cluster 1 is further partitioned into two subclasses: large-scale foundation-model training (Subclass 1a, Frontier) and smaller-scale training and batch tasks (Subclass 1b, Batch) [6]. This section develops the mathematical model of the AIDC as a set of physical and operational constraints.

### *A. Cluster-Level Modeling Elements*

Let $k \in \{1a, 1b, 2\}$ index the three workload groups. Each group is composed of $N_k$ homogeneous GPU accelerators of identical type, and is controlled at the group level rather than at the individual accelerator level.

***Throughput.*** The operational state of group $k$ at discrete time step $t \in \{1,2,\dots,T\}$ is characterized by a continuous variable $s_k(t) \in [0,1]$ , representing the normalized computational throughput as a fraction of peak capacity:

$$0 \le s_k(t) \le 1, k \in \{1a, 1b, 2\} \tag{2}$$

when $s_k(t) = 0$, the cluster operates in deep idle, all accelerators remain powered on at their idle baseline but perform no useful computation. Full shutdown (power-off) is not modeled, reflecting the operational reality of modern GPU clusters in which voltage regulators, high-bandwidth memory, and interconnect fabrics remain energized at all times.

***Group IT power consumption.*** Empirical measurements of modern AI accelerators show that GPU electrical power varies approximately linearly with normalized throughput within the operational range [21, 22]. Accordingly, the aggregate IT power consumption of group k is modelled as:

$$P_{\mathrm{IT},k}(t) = P_k^{\mathrm{idle}} + \left(P_k^{\mathrm{peak}} - P_k^{\mathrm{idle}}\right) \cdot s_k(t) \tag{3}$$

where $P_k^{\mathrm{peak}}$ is the aggregate peak power of group $k$ (when $s_k(t) = 1$), and $P_k^{\mathrm{idle}}$ is the aggregate idle power (when $s_k(t) = 0$).

***Total IT power.*** The total IT power consumed by the AIDC is the sum across groups:

$$P_{\mathrm{IT}}(t) = \sum_{k \in \{1a, 1b, 2\}} P_{\mathrm{IT},k}(t) \tag{4}$$

### *B. Training Cluster 1: Frontier and Batch Cluster*

Cluster 1 hosts training workloads that are assigned cumulative throughput targets over the scheduling horizon. The two subclasses (Frontier and Batch) differ not in their mathematical constraint structure but in their operational characteristics and the economic consequences of under-delivery. Subclass 1a (Frontier) hosts large-scale foundation-model pre-training workloads. These runs typically span days to weeks [23], and any shortfall in delivered throughput delays milestone completion with high economic and reputational cost. Subclass 1b (Batch) hosts smaller-scale training workloads such as fine-tuning of pre-trained models, research experimental runs, batch inference jobs, and offline data preprocessing pipelines. These workloads are short-lived (minutes to hours), and job-level resubmission is routine [6].

***Workload delivery.*** Each training subclass $k \in \{1a, 1b\}$ is assigned a cumulative workload target $w_{\mathrm{req},k}$ to be completed by the end of the scheduling horizon. Under-delivery is represented by a nonnegative slack:

$$\sum_{t=1}^{T} s_k(t) \cdot \Delta t \ge w_{\mathrm{req},k} - S_{\mathrm{W},k}, S_{\mathrm{W},k} \ge 0 \tag{5}$$

The slack $S_{\mathrm{W},k}$ enters the operational objective (Section V) with a group-specific penalty weight $M_k$. The key modeling distinction between the two subclasses is encoded entirely through this penalty: $M_{1a} \gg M_{1b}$, reflecting the fact that frontier training under-delivery is far more costly than batch under-delivery. This asymmetry, rather than any difference in physical constraints, is what produces the differentiated flexibility profiles observed in the case studies of Section VI; the planning layer preferentially curtails batch throughput during grid stress while protecting frontier training progress. At the 15-minute control timescale of this paper, GPU clusters can change their computational throughput via dynamic voltage-frequency scaling within seconds, so no inter-step ramping or inertia constraint is imposed.

### C. Cluster 2: Inference Cluster

Cluster 2 hosts customer-facing inference services. Unlike training, inference workload arrives exogenously from external users at each time step and is not accumulated as a deferrable backlog. The constraint set below captures the partial-serving capability of modern inference orchestration systems [24].

***Exogenous demand and partial serving.*** Let $d_{\text{inf}}(t) \in [0,1]$ denote the normalized inference request arrival rate at time $t$, expressed relative to the peak processing capacity of Cluster 2. The throughput variable $s_2(t)$ represents the served portion of incoming requests, bounded above by the request arrival rate:

$$s_2(t) \le d_{\text{inf}}(t) \tag{6}$$

The rejected request rate at time t is:

$$r_2(t) = d_{\text{inf}}(t) - s_2(t) \tag{7}$$

when $s_2(t) = 0$ (deep idle), $r_2(t) = d_{\text{inf}}(t)$, reflecting that all incoming requests are rejected. When $s_2(t) = d_{\text{inf}}(t)$, all requests are served and $r_2(t) = 0$.

### D. Cooling Subsystem

The cooling overhead of a gigawatt-scale AIDC is characterized by its power usage effectiveness (PUE), defined as the ratio of total facility power to IT equipment power. State-of-the-art facilities employing direct-to-chip liquid cooling (DLC) achieve PUE values near 1.10, indicating that cooling and power distribution losses add approximately 10% to the IT load. The cooling electrical power consumption is accordingly modeled as a fixed fraction of total IT power:

$$P_{\text{cool}}(t) = \gamma \cdot P_{\text{IT}}(t) \tag{8}$$

where $\gamma = \text{PUE} - 1$. This linear relationship is a standard approximation in data center energy modelling [25] and is particularly accurate for DLC-cooled facilities, where the cooling overhead is small and largely determined by fixed infrastructure (pumps, coolant distribution units, and heat rejection systems) rather than by dynamic thermal control. At the 15-minute control timescale of this paper, γ is treated as a constant; seasonal or diurnal variations in PUE due to ambient temperature changes are neglected.

### E. Battery Energy Storage System

The on-site BESS serves as a power buffer between the PCC-accepted power and the AIDC's internal consumption. The BESS dispatch is described by a binary mode indicator $\beta(t) \in \{0,1\}$ together with two nonnegative continuous variables $P_{\text{ch}}(t)$ and $P_{\text{dis}}(t)$ representing charging and discharging power, respectively. The net BESS power is:

$$P_{\text{bess}}(t) = P_{\text{ch}}(t) - P_{\text{dis}}(t) \tag{9}$$

where $P_{\text{bess}}(t) > 0$ indicates charging (additional power drawn from the grid) and $P_{\text{bess}}(t) < 0$ indicates discharging. The charging and discharging constraints:

$$0 \le P_{\text{ch}}(t) \le \beta(t) P_{\text{bess}}^{\max}, 0 \le P_{\text{dis}}(t) \le (1-\beta(t)) P_{\text{bess}}^{\max} \tag{10}$$

The state of charge (SoC) evolves according to:

$$E_{\text{bess}}(t+1) = E_{\text{bess}}(t) + \left[\eta_{\text{ch}} \cdot P_{\text{ch}}(t) - \frac{P_{\text{dis}}(t)}{\eta_{\text{dis}}}\right] \cdot \Delta t \tag{11}$$

where $\eta_{\text{ch}}, \eta_{\text{dis}} \in [0,1]$ are charging and discharging efficiencies. The SoC is bounded:

$$E_{\text{bess}}^{\min} \le E_{\text{bess}}(t) \le E_{\text{bess}}^{\max} \tag{12}$$

A cyclic terminal condition is imposed to maintain a well-defined long-horizon evaluation:

$$E_{\text{bess}}(T+1) = E_{\text{bess}}(1) = 0.9 \cdot E_{\text{bess}}^{\max} \tag{13}$$

The BESS is initialized at high state of charge at the beginning of each scheduling episode, reflecting the operational practice of maintaining on-site storage in a near-full standby state.

### F. Grid-Interfaced Power Balance and AIDC Feasibility Region

The electrical power drawn by the AIDC from the PCC must equal the sum of all internal loads after accounting for power conditioning losses. As described above, the DC-path components (IT load and BESS) pass through the IPCS with efficiency $\eta_{\text{IPCS}}$, while the AC-path cooling load is drawn directly. The AIDC power balance is:

$$P_{\text{acc}}(t) = \frac{P_{\text{IT}}(t) + P_{\text{bess}}(t)}{\eta_{\text{IPCS}}} + P_{\text{cool}}(t), P_{\text{acc}}(t) \ge 0 \tag{14}$$

where $P_{\text{acc}}(t)$ is the accepted PCC import and $\eta_{\text{IPCS}} \in (0,1)$ is the aggregate efficiency of the internal power conditioning system. Substituting (4) and (8), the grid-side demand separates into a DC-path component $\frac{P_{\text{IT}}(t)+P_{\text{bess}}(t)}{\eta_{\text{IPCS}}}$ and an AC-path cooling component $\gamma P_{\text{IT}}(t)$. The non-negativity of $P_{\text{acc}}(t)$ reflects the consumption-only nature of the AIDC at the grid interface.

***AIDC feasibility region.*** Collecting all AIDC decision variables at time t as: $x(t) = \{s_k(t), P_{\text{ch}}(t), P_{\text{dis}}(t)\}$, for $k \in \{1a, 1b, 2\}$, and given the accepted PCC exchange $P_{\text{acc}}(t)$ together with the battery state $E_{\text{bess}}(t)$ at time $t$, the AIDC feasibility region is defined as:

$$\mathcal{F}_{\text{AIDC}}\big(P_{\text{acc}}(t), E_{\text{bess}}(t)\big) = \left\{ \begin{matrix} x(t) \\ \text{s.t. } (2)-(4), (6)\text{–}(7), (9)-(14) \end{matrix} \right\}. \tag{15}$$

The mapping $\mathcal{F}_{\text{AIDC}}(\cdot)$ characterizes the set of physically admissible internal operating decisions compatible with a given accepted PCC exchange at time $t$. The feasibility region is non-empty for any $P_{\text{acc}}(t) \ge P_{\text{idle}}^{\min}$, where $P_{\text{idle}}^{\min} = \left(\frac{1}{\eta_{\text{IPCS}}} + \gamma\right) \sum_k P_k^{\text{idle}}$ is the aggregate idle power draw: idling the BESS ($P_{\text{ch}} = P_{\text{dis}} = 0$) yields a feasible deep-idle operating point. This lower bound is always satisfied in practice, as the AIDC's minimum power request never falls below $P_{\text{idle}}^{\min}$ and the TSO acceptance mechanism constrains curtailment to $\kappa \le P_{\text{req}}$.

## IV. Transmission Network Feasibility Model

This section presents the transmission network feasibility model used to characterize the set of physically admissible grid operating conditions under a given AIDC power exchange at the PCC. The network is represented via a linearized DC power flow with generator operating envelopes, ramping constraints, and a budget-constrained uncertainty set modeling short-term demand deviations. The transmission system is assumed to operate securely and feasibly in the absence of the AIDC.

### A. Nodal Power Balance and Network Representation

The transmission network is represented by a set of buses $\mathcal{N}$, a set of transmission lines $\mathcal{L}$, and a set of generators $\mathcal{G}$. Each generator $i \in \mathcal{G}$ is connected at a designated bus $n_i \in \mathcal{N}$. The AIDC is connected at a designated bus $n_a \in \mathcal{N}$. A linearized DC power flow representation is adopted, as the connect-and-manage curtailment decisions considered in this paper are specified in active-power terms and determined primarily by thermal congestion and generator dispatch limits.

***Nodal active power balance.*** At each bus $n \in \mathcal{N}$ and time $t$:

$$\sum_{i:n_i=n} g_i(t) - d_n(t,\boldsymbol{\xi}) - \mathbb{1}_{n=n_a} P_{\text{acc}}(t) = \sum_{l\in\delta^+(n)} f_l(t) - \sum_{l\in\delta^-(n)} f_l(t) \quad (16)$$

where $g_i(t)$ is the active power output of generator $i$; $d_n(t,\boldsymbol{\xi})$ is the realized background demand at bus $n$ under uncertainty realization $\boldsymbol{\xi}$ (formalized in IV.C); $\mathbb{1}_{n=n_a}$ is an indicator equal to 1 if $n = n_a$ and 0 otherwise; $f_l(t)$ is the active power flow on line $l$; and $\delta^+(n), \delta^-(n)$ denote the sets of lines entering and leaving bus $n$, respectively.

***Line flow representation.*** For each line $l = (i,j) \in \mathcal{L}$, the power flow is determined by the phase-angle difference and line susceptance:

$$f_l(t) = B_l \cdot \left(\theta_i(t) - \theta_j(t)\right) \quad (17)$$

where $\theta_n(t)$ is the voltage phase angle at bus $n$ and $B_l$ is the line susceptance. The phase angle at a designated reference bus is fixed to zero to ensure uniqueness.

***Line thermal limits.*** The active power flow on each transmission line is bounded by its thermal capacity:

$$-F_l^{\max} \le f_l(t) \le F_l^{\max} \quad (18)$$

where $F_l^{\max}$is the thermal rating of line $l$.

## B. Generator Operating Constraints

The TSO's ability to determine generator dispatch $\{g_i(t)\}$ in this formulation reflects standard institutional arrangements in modern wholesale electricity markets. In market-based systems, the TSO does not own generation assets but exercises operational dispatch authority through security-constrained market clearing: generators submit offers and are economically obligated to follow the resulting dispatch instructions.

***Generator output bounds.*** Each generator's active power output is bounded:

$$g_i^{\min} \le g_i(t) \le g_i^{\max}, \forall i \in \mathcal{G} \quad (19)$$

where $g_i^{\min}$ and $g_i^{\max}$ are the minimum and maximum output capacities.

***Generator ramp constraints.*** To reflect the physical limitation that conventional generators cannot change output arbitrarily fast, a ramp-rate constraint is imposed between consecutive time steps:

$$|g_i(t) - g_i(t-1)| \le R_i^{\text{gen}} \cdot \Delta t, \forall i \in \mathcal{G} \quad (20)$$

where $R_i^{\text{gen}}$ is the maximum ramp rate of generator $i$. The ramp-rate parameter is differentiated by generator technology in this work: baseload units (nuclear, coal) have low ramp rates, mid-merit gas combined-cycle units have moderate rates, and gas peakers have high rates, following standard industry classifications.

## C. Demand Uncertainty via Budget Uncertainty Set

Short-term background demand is forecast with non-trivial error. The TSO's feasibility model must accommodate a range of admissible demand realizations, rather than only the point forecast. To this end, the demand uncertainty is represented as a budget uncertainty set [26].

***Uncertainty parameterization.*** Let $\hat{d}_n(t)$ denote the point forecast of demand at bus $n$at time $t$. The realized demand under uncertainty is:

$$d_n(t,\boldsymbol{\xi}) = \hat{d}_n(t) + \xi_n(t) \cdot \hat{d}_n(t) \cdot \epsilon, \forall n \in \mathcal{N} \quad (21)$$

where $\xi_n(t) \in [-1,1]$ is a normalized perturbation variable for bus $n$ at time $t$, and $\epsilon \in (0,1)$ is the maximum single-node deviation ratio relative to forecast.

***Uncertainty set.*** The budget uncertainty set is:

$$\mathcal{U} = \{\boldsymbol{\xi}: |\xi_n(t)| \le 1, \forall n,t; \textstyle\sum_{n\in\mathcal{N}} |\xi_n(t)| \le \Gamma_{\text{U}}, \forall t\} \quad (22)$$

where $\Gamma_{\text{U}} \in [0, |\mathcal{N}|]$ is the budget parameter bounding the number of buses whose demand may simultaneously deviate to the maximum extent. When $\Gamma_{\text{U}} = 0$, the set collapses to the forecast. When $\Gamma_{\text{U}} = |\mathcal{N}|$, the set coincides with the worst-case box uncertainty.

## D. Baseline Dispatch Reference

In this paper, $\{g_i^0(t)\}_{t\in\mathcal{T}}$ is computed by solving a single AIDC-free deterministic DC optimal power flow (DC-OPF) over the entire horizon: minimize $\sum_{t\in\mathcal{T}} \sum_{i\in\mathcal{G}} c_i \cdot g_i(t)$ subject to the nodal balance (16), line flow (17), line thermal limits (18), generator output bounds (19), and generator ramp constraints (20) coupling consecutive time steps, with $d_n(t) = \hat{d}_n(t)$ (point forecast) and $P_{\text{acc}}(t) = 0$ (no AIDC). This approximates the day-ahead market-cleared schedule in the absence of real-time demand uncertainty or strategic behavior, and ensures that the baseline trajectory itself is ramp-feasible. More sophisticated day-ahead clearing mechanisms can be substituted without altering the formulation in Section V. The baseline $\{g_i^0(t)\}$ is pre-computed before the AIDC–TSO interaction begins and treated as an exogenous parameter throughout.

## E. Transmission Feasibility

Given the constraints developed in IV.A–IV.C, the transmission feasibility region for a candidate accepted exchange $P_{\text{acc}}(t)$ and uncertainty realization $\boldsymbol{\xi}$ is:

$$\mathcal{F}_{\text{TSO}}(P_{\text{acc}}(t), \boldsymbol{\xi}) = \left\{ \begin{matrix} \{g_i(t)\}_{i\in\mathcal{G}}, \{\theta_n(t)\}_{n\in\mathcal{N}}, \{f_l(t)\}_{l\in\mathcal{L}} \\ \text{s.t. } (16)-(20) \text{ with } d_n(t,\boldsymbol{\xi}) \text{ per } (21) \end{matrix} \right\}. \quad (23)$$

The mapping $\mathcal{F}_{\text{TSO}}(\cdot)$ characterizes the set of physically admissible generator dispatches, voltage angles, and line flows compatible with a given accepted AIDC exchange under a given demand realization. A candidate exchange is said to be robustly feasible if $\mathcal{F}_{\text{TSO}}(P_{\text{acc}}(t), \boldsymbol{\xi}) \ne \emptyset$ for all $\boldsymbol{\xi} \in \mathcal{U}$. The baseline dispatch $g_i^0(t)$ defined in IV.D provides a reference configuration within $\mathcal{F}_{\text{TSO}}$ under AIDC-free conditions ($P_{\text{acc}}(t) = 0$).

# V. Hierarchical Decision Problem under Connect-and-Manage

The feasibility regions $\mathcal{F}_{\text{AIDC}}$ and $\mathcal{F}_{\text{TSO}}$ characterize the physical admissibility of AIDC and grid operating decisions. This section develops the decision-making structure that operates on these regions under the interaction protocol of Section II. The formulation is organized into three layers: the TSO determines the accepted PCC exchange through security-constrained robust optimization; the AIDC maintains a capacity-aware planning policy that sets operational targets in anticipation of TSO behavior; and an execution optimizer translates the planning targets into physically feasible operating decisions at each time step.

## A. TSO Robust Power Acceptance Problem

At each time step t, upon receiving $P_{\text{req}}(t)$, the TSO solves a robust optimization problem over $\mathcal{F}_{\text{TSO}}$. The decision variables are the generator dispatch $\{g_i(t)\}$, voltage angles $\{\theta_n(t)\}$, line flows $\{f_l(t)\}$, and curtailment $\kappa(t)$. The generator ramp constraint (20) couples $g_i(t)$ to the previous-step realized dispatch $g_i(t-1)$. The TSO minimizes:

$$\min \Gamma_\kappa \cdot \kappa(t) + \sum_{i\in\mathcal{G}} c_i \cdot g_i(t) + \rho \cdot \sum_{i\in\mathcal{G}} \left|g_i(t) - g_i^0(t)\right| \quad (24)$$

where $\Gamma_\kappa > 0$ penalizes curtailment, $c_i$ is the marginal cost of generator i, and $\rho > 0$ penalizes deviation from the baseline dispatch. The weights satisfy:

$$\Gamma_\kappa \gg \max_i c_i \gg \rho \quad (25)$$

This ordering encodes the TSO's institutional role as a non-profit security authority: curtailment is a last-resort physical enforcement mechanism, invoked only when re-dispatch within the feasible region cannot satisfy network constraints. When constraints are slack, the optimal solution yields $\kappa(t) = 0$ and $P_{\text{acc}}(t) = P_{\text{req}}(t)$.

***Constraints.*** Subject to the coupling (1), the transmission feasibility constraints (16)–(20), the demand realization (21), and robust feasibility: the solution must satisfy (16)–(18) with $d_n(t, \boldsymbol{\xi})$ for every $\boldsymbol{\xi} \in \mathcal{U}$.

***PCC admission ramp limit.*** The upward rate of change of the accepted exchange is bounded:

$$|P_{\text{acc}}(t) - P_{\text{acc}}(t-1)| \le R_{\text{grid}} \quad (26)$$

where $R_{\text{grid}}$ is the maximum admissible step at the PCC.

***Acceptance mapping.*** Under the budget uncertainty set (22) and the linear structure of (16)–(20), the robust problem admits a tractable dual reformulation and is solved as an LP [26]. The optimal solution induces an implicit acceptance mapping:

$$P_{\text{acc}}(t) = F_{\text{TSO}}(P_{\text{req}}(t); g^0(t), g(t-1), \mathcal{U}, \text{network data}) \quad (27)$$

which is the black-box function the AIDC learns to anticipate. Curtailment arises as a derived outcome of physical constraint binding through three structurally distinct mechanisms: congestion-binding, ramp-binding, and robustness-binding. These mechanisms exhibit distinct temporal signatures and are analyzed empirically in Section VI.

*B. AIDC Capacity-Aware Planning Policy*

Under the connect-and-manage protocol, the AIDC's strategic problem is how to organize its operation over time, namely workload pacing across clusters, inference utilization, such that delivery commitments are preserved under expected curtailment patterns. Because $\mathcal{F}_{\text{TSO}}(\cdot)$ is not disclosed, the policy is learned through repeated interaction with the TSO. The planning problem is formulated as a Markov decision process (MDP) over the scheduling horizon.

***State.*** The state at time $t$ is:

$$o_t = \{E_{\text{bess}}(t), \{s_k(t-1)\}_k, R_{1a}(t), R_{1b}(t), \eta_{1a}(t), \eta_{1b}(t), P_{\text{acc}}(t-1), \kappa(t-1), \pi(t), D(t), d_{\text{inf}}(t)\} \quad (28)$$

where $E_{\text{bess}}(t)$ is the BESS SoC; $\{s_k(t-1)\}_k$ are the prior-step cluster throughputs; $R_{1a}(t)$, $R_{1b}(t)$ are the remaining cumulative workload targets for the two training subclasses; and the urgency indicators

$$\eta_k(t) = \frac{R_k(t)\cdot T}{w_{\text{req},k}\cdot(T-t+1)} \quad \text{for } k \in \{1a, 1b\} \quad (29)$$

yield a dimensionless urgency ratio. Values $\eta_k(t) < 1$ indicate the AIDC is ahead of schedule; $\eta_k(t) > 1$ indicates the AIDC must accelerate to complete the target workload.

***Action.*** The action is a five-dimensional continuous planning vector:

$$a_t = (\tilde{s}_{1a}, \tilde{s}_{1b}, \tilde{s}_2, \tilde{\varphi}_{\text{ch}}, \tilde{\varphi}_{\text{dis}})_t \in [0,1]^5 \quad (30)$$

representing the target throughput for each cluster and the normalized BESS charging and discharging power. The continuous action space is handled directly by the soft actor-critic (SAC) reparameterization trick [27]. Given the planning action, the requested PCC exchange is computed from the internal power balance:

$$P_{\text{req}}(t) = \left(\frac{1}{\eta_{\text{IPCS}}} + \gamma\right) \sum_k \left(P_k^{\text{idle}} + \left(P_k^{\text{peak}} - P_k^{\text{idle}}\right) \tilde{s}_k(t)\right) + \frac{1}{\eta_{\text{IPCS}}} \left(\tilde{P}_{\text{ch}}(t) - \tilde{P}_{\text{dis}}(t)\right) \quad (31)$$

The request reflects the aggregate power required to execute the planning intent, inclusive of IT load, cooling, IPCS losses, and net BESS power exchange ( $\tilde{P}_{\text{ch}}(t) = \tilde{\varphi}_{\text{ch}} P_{\text{bess}}^{\max}$ , $\tilde{P}_{\text{dis}}(t) = \tilde{\varphi}_{\text{dis}} P_{\text{bess}}^{\max}$).

***State transition.*** Given the action, the environment transition proceeds as follows: $P_{\text{req}}(t)$ is computed via (31) and submitted to the TSO; the TSO returns $P_{\text{acc}}(t)$ via (27); the execution optimizer (V.C) determines the actual operation and the internal state updates.

***Reward.*** The per-step reward captures workload delivery pressure and curtailment avoidance:

$$r_t = -\alpha_w\left(M_{1a}\hat{\eta}_{1a}(t) + M_{1b}\hat{\eta}_{1b}(t)\right) - \alpha_{\text{rej}} r_2(t)\Delta t - \alpha_\kappa \kappa(t) \quad (32)$$

where the normalized workload shortfall ratio is: $\hat{\eta}_k(t) = \max\left(0, \frac{t}{T} w_{\text{req},k} - W_k(t)\right) / w_{\text{req},k}$, $W_k(t) = \sum_{\tau=1}^{t} s_k(\tau) \cdot \Delta t$. The penalty weights satisfy $M_{1a} \gg M_{1b}$, encoding the asymmetric cost of under-delivery between Frontier and Batch training. The three terms in (32) represent: (i) workload urgency penalty, which increases when cumulative throughput falls behind the uniform delivery schedule; (ii) inference rejection cost, proportional to the unserved request rate $r_2(t)$ defined in (7); and (iii) curtailment penalty, providing a direct incentive for the planning policy to anticipate and avoid TSO curtailment.The policy $\pi_\theta$ maximizes the expected cumulative return:

$$\max_\theta \mathbb{E}\left[\sum_{t=1}^{T} r_t\right] \quad (33)$$

*C. Execution Optimizer*

Given the planning action $a_t$ and the realized accepted exchange $P_{\text{acc}}(t)$, the execution optimizer determines the AIDC's internal operating decisions by solving a single-step mixed-integer linear program (MILP) over $\mathcal{F}_{\text{AIDC}}\left(P_{\text{acc}}(t), E_{\text{bess}}(t)\right)$. The optimizer serves three roles: (i) it maps the continuous planning targets to physically feasible cluster throughputs under the binding power budget; (ii) it determines the BESS dispatch that satisfies the exact power balance (14); and (iii) when $P_{\text{acc}}(t) < P_{\text{req}}(t)$, it reallocates the reduced power budget across clusters and BESS to minimize operational cost. At time t, the optimizer determines: $\chi(t) = \{s_{1a}(t), s_{1b}(t), s_2(t), P_{\text{ch}}(t), P_{\text{dis}}(t)\}$

***Objective.*** The execution objective combines L1 tracking penalty with BESS cycling cost on throughput deviation from the planning targets:

$$\min \lambda \sum_k u_k(t) + C_{\text{cyc}}\left(P_{\text{ch}}(t) + P_{\text{dis}}(t)\right)\Delta t \quad (34)$$

where the tracking auxiliaries enforce:

$$u_k(t) \ge s_k(t) - \tilde{s}_k(t), u_k(t) \ge \tilde{s}_k(t) - s_k(t), u_k(t) \ge 0 \quad (35)$$

where the cycling coefficient $C_{\text{cyc}}$ set sufficiently low that the optimizer prefers discharging the BESS over reducing cluster throughput when curtailment occurs, enabling the BESS to serve as a power buffer that preserves workload delivery. The tracking weight $\lambda$ ensures execution follows the planning intent when $P_{\text{acc}}(t) = P_{\text{req}}(t)$, but yields to physical feasibility when $P_{\text{acc}}(t) < P_{\text{req}}(t)$.

*D. Closed-Loop Training and Execution*

The three layers interact in closed loop through repeated

AIDC–TSO interaction. During offline training, the closed loop is simulated over historical environment trajectories, allowing the policy $\pi_\theta$ to converge via SAC updates from observed transitions. After training, the policy parameters $\theta$ are fixed for online execution. Algorithms 1 and 2 summarize the offline training and online execution procedures, respectively.

***Algorithm 1:*** *Offline Training of the Capacity-Aware Planning Policy*

**Input:** Environment parameters (AIDC, BESS, network); baseline dispatch $\{g_i^0(t)\}$; exogenous signal trajectories $\{\pi(t), D(t), d_{\text{inf}}(t), \hat{d}_n(t)\}$; SAC hyperparameters; training episodes $N_{\text{ep}}$.
**Output:** Trained request policy $\pi_\theta(\cdot)$.

1. Initialize policy $\pi_\theta$, Q-networks, and replay buffer $\mathcal{D}$.
2. **for episode** $= \mathbf{1, 2, \ldots}, N_{\text{ep}}$ **do**
3. Initialize internal state $(E_{\text{bess}}(1), \{s_k(0)\}, R_k(1), g(0))$; sample exogenous trajectories over the horizon.
4. **for** $\boldsymbol{t = 1, 2, \ldots, T}$ **do**
5. Construct state $o_t$ via (28)–(29).
6. Sample planning action $a_t \sim \pi_\theta(\cdot \mid s_t)$; compute $\tilde{P}_{\text{ch}}(t)$, $\tilde{P}_{\text{dis}}(t)$ via (30); compute $P_{\text{req}}(t)$ via (31).
7. **[TSO]** Solve robust acceptance problem (1), (24)–(27) to obtain $P_{\text{acc}}(t), \kappa(t), g(t)$.
8. **[AIDC]** Solve execution optimizer (34)–(35) over $\mathcal{F}_{\text{AIDC}}(P_{\text{acc}}(t), E_{\text{bess}}(t))$.
9. Compute per-step reward $r_t$ via (32); update internal state $E_{\text{bess}}(t)$ via (11), $R_k(t+1) \leftarrow R_k(t) - s_k(t)\Delta t$.
10. Store transition $(o_t, a_t, r_t, o_{t+1})$ in $\mathcal{D}$; perform SAC update on $\pi_\theta$ and Q-networks.
11. **end for**
12. **end for**
13. **return** trained policy $\pi_\theta$.

***Algorithm 2:*** *Online Closed-Loop AIDC–TSO Operation*

**Input:** Trained planning policy $\pi_\theta$ (from Algorithm 1); pre-computed baseline dispatch $\{g_i^0(t)\}$; initial internal state $(E_{\text{bess}}(1), \{s_k(0)\}, R_k(1))$.
**Output:** Executed trajectories $\{x(t)\}$; interaction record $\{P_{\text{req}}(t), P_{\text{acc}}(t), \kappa(t)\}$; terminal workload delivery $\{W_k(T)\}$.

1. **for** $\boldsymbol{t = 1, 2, \ldots, T}$ **do**
2. Observe exogenous signals $\pi(t), D(t), d_{\text{inf}}(t)$; construct state $o_t$ via (28)–(29).
3. Execute planning policy $a_t \leftarrow \pi_\theta(s_t)$ (deterministic); compute $\tilde{P}_{\text{ch}}(t)$, $\tilde{P}_{\text{dis}}(t)$ via (30); compute $P_{\text{req}}(t)$ via (31).
4. **[TSO]** Solve robust acceptance problem (1), (24)–(27) → $P_{\text{acc}}(t), \kappa(t)$.
5. **[AIDC]** Solve execution optimizer (34)–(35) over $\mathcal{F}_{\text{AIDC}}(P_{\text{acc}}(t), E_{\text{bess}}(t))$.
6. Implement $x(t)$; update internal state: $E_{\text{bess}}(t)$ via (11), $R_k(t+1) \leftarrow R_k(t) - s_k(t)\Delta t$; record interaction outcome.
7. **end for**
8. Compute terminal workload delivery $W_k(T) = \sum_{\tau=1}^{T} s_k(t)\Delta t$ for $k \in \{1a, 1b\}$.
9. **return** $\{x(t)\}, \{P_{\text{req}}(t), P_{\text{acc}}(t), \kappa(t)\}, \{W_k(T)\}$.

## VI. Case Studies

### A. Test System, Data, and Implementation

The IEEE 39-bus system is selected because its concentrated generation portfolio and limited transmission corridors create operationally relevant congestion under gigawatt-scale AIDC loading. Larger test systems with more distributed generation and denser transmission networks would dilute the congestion effect and reduce curtailment frequency, making the interaction protocol less relevant in the case study. The AIDC is connected at Bus 16, a five-branch hub node whose central location exposes the facility to congestion on multiple corridors. Line thermal ratings are scaled to 78% of nominal values so that the default TSO robustness configuration ( $\Gamma_{\text{U}}$ =5, $\varepsilon$ =0.07) produces a curtailment frequency of approximately 3%, representative of the curtailment levels contemplated in current connect-and-manage proposals. Generator marginal costs range from 10 AUD/MWh (nuclear) to 120 AUD/MWh (gas peaker), and ramp rates are differentiated by technology type (80–1,200 MW/h). All system and algorithm parameters are listed in Table I. The AIDC has a total IT capacity of 1,100 MW with peak grid-side demand of approximately 1,268 MW. The workload targets for Frontier and Batch training are both set to 94.0% of peak throughput, representing the maximum achievable delivery rate after accounting for curtailment-induced throughput losses under Always Full dispatch. Exogenous data are drawn from the AEMO New South Wales region at 15-minute resolution. The training set covers August 2023 through January 2024 (6 months), and the test set covers the entirety of February 2024, ensuring zero temporal overlap between training and evaluation periods. The SAC planning policy is trained using Stable-Baselines3 (SB3) with hyperparameters listed in Table I. Deep reinforcement learning (DRL) baselines (DDPG, TD3) are trained under identical environment and reward configurations.

TABLE I System Parameters

| | Parameter | Value |
|---|---|---|
| Grid | $F_l^{\max}$ | 78% of nominal |
| | $c_i$ | 10–120 AUD/MWh |
| | $R_i^{\text{gen}}$ | 80–1,200 MW/h |
| | $\Gamma_{\text{U}} / \varepsilon$ | 5 / 0.07 |
| | $\Gamma_\kappa$ | $10^5$ |
| | $\rho$ | 1 |
| | $R_{\text{grid}}$ | 150 MW/step |
| AIDC | $P_k^{\text{peak}}$ | 550 / 220 / 330 MW |
| | $P_k^{\text{idle}} / P_k^{\text{peak}}$ | 0.30 / 0.25 / 0.20 |
| | $\gamma / \eta_{\text{IPCS}}$ | 0.10 / 0.95 |
| | $w_{\text{req},k}$ | 94.0% |
| BESS | $P_{\text{bess}}^{\max}$ | 200 MW |
| | $E_{\text{bess}}^{\min} / E_{\text{bess}}^{\max}$ | 30MWh/300MWh |
| | $\eta_{\text{ch}}, \eta_{\text{dis}}$ | 0.95 / 0.95 |
| | $SoC_{\text{init}}$ | 0.9 |
| | $C_{\text{cyc}}$ | 0.5 |
| Reward | $\lambda$ | 100 |
| | $\alpha_w$ | 0.01 |
| | $M_{1a}/M_{1b}$ | 100 / 50 |
| | $\alpha_\kappa$ | 0.005 |
| | $\alpha_{\text{rej}}$ | 3 |
| SAC | Network | MLP [256, 256] |
| | Learning rate | $3 \times 10^{-4}$ |
| | Replay buffer | $5 \times 10^5$ |
| | Batch / Discount | 256 / 0.99 |

All experiments are conducted on a laptop with an Intel Core Ultra 9 275HX processor and an NVIDIA RTX 5090 GPU. The TSO acceptance and AIDC execution optimization problems are solved using Gurobi 11.0. The RL algorithms are implemented with PyTorch.

The case studies in this section validate the proposed framework. Section VI.B compares power request strategies, evaluating the proposed learning-based approach against rule-based and heuristic alternatives. Section VI.C examines the differentiated cluster scheduling produced by the planning layer under grid stress. Section VI.D analyzes how the execution optimizer and the on-site BESS jointly respond to residual curtailment events. Section VI.E traces the propagation of curtailment intensity to AIDC workload delivery across a range of TSO operating conditions.

### B. AIDC Power Request Strategy Comparison

Five power request strategies are compared on the held-out evaluation period: a conservative static margin (Fixed-Buffer-85), a rule-based demand-threshold policy (Heuristic), and three DRL algorithms (DDPG, TD3, SAC). All strategies use the same execution optimizer; only the planning-layer logic differs. The results are summarized in Table II.

TABLE II: AIDC POWER REQUEST STRATEGY COMPARISON

| Strategy | Reward | Mean $\kappa$ (MW) | Curt % | $W_{1a}$ | $W_{1b}$ |
|---|---|---|---|---|---|
| Fixed-Buffer-85 | −161.8 | 22.79 | 9.1% | 89.4% | 90.8% |
| Heuristic | −74.5 | 9.60 | 3.9% | 94.6% | 88.3% |
| DDPG | −128.8 | 3.50 | 5.4% | 91.3% | 56.5% |
| TD3 | −60.1 | 2.28 | 2.9% | 94.0% | 86.2% |
| **SAC** | **−29.3** | **1.03** | **2.8%** | **98.1%** | **88.9%** |

Fixed-Buffer-85 achieves the highest curtailment frequency (9.1%) and the lowest Frontier workload completion ($W_{1a}$= 89.4%) among all five strategies. The Heuristic reduces curtailment to 3.9% while maintaining substantially higher workload completion ($W_{1a}$=94.6%, $W_{1b}$=88.3%), achieving the strongest non-learning performance (reward −74.5). The contrast between these two strategies illustrates that demand-responsive scheduling significantly outperforms static margins.

SAC achieves the best overall performance: the lowest mean curtailment (1.03 MW), the highest Frontier completion ($W_{1a}$= 98.1%), and the best reward (−29.3), improving over the Heuristic by 60.7%. Notably, SAC and TD3 achieve almost the same curtailment frequency (2.8% and 2.9%), but SAC delivers substantially higher workload ($W_{1a}$=98.1% vs 94.0%) and better reward (−29.3 vs −60.1), indicating that SAC uses the curtailment headroom more effectively to maximize workload throughput.

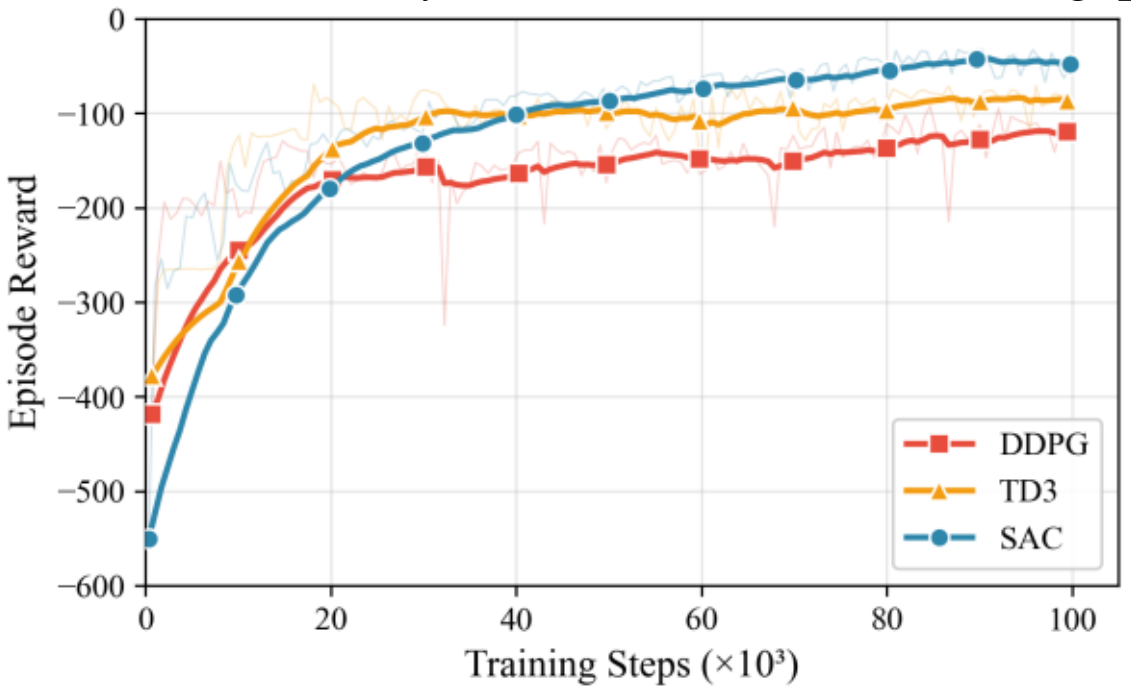

Fig. 2 Training convergence of the three DRL algorithms.

Fig. 2 compares the training convergence of DDPG, TD3, and SAC under identical environment configurations, network architectures, and reward functions. DDPG converges the slowest and plateaus at the worst reward level, consistent with its poor workload delivery in Table II. Its single-critic architecture is prone to Q-value overestimation, leading to unstable request policies that fail to consistently avoid curtailment while also under-delivering workload. TD3 achieves noticeably faster initial convergence and a better final reward than DDPG, benefiting from twin critics that mitigate overestimation. SAC converges the fastest and achieves the best final reward. The entropy-regularized stochastic policy enables broader exploration of the request space, which is essential when the TSO acceptance mapping is opaque to the AIDC. All three algorithms exhibit increasing stability in the later training phase, but only SAC consistently reaches the reward region associated with effective curtailment avoidance in Table II.

### C. Heterogeneous Flexibility Under Grid Stress

Fig. 3 shows the weekly operation of the SAC policy on the evaluation period. The upper panel displays the power request and accepted exchange at the PCC. Throughout the week, $P_{req}$ fluctuates between 600 and 1,100 MW, reaching approximately 1,200 MW during off-peak periods on February 23, and dropping as low as 700 MW and 600 MW during the most severe grid stress events on February 23 and February 29. The accepted exchange $P_{\mathrm{acc}}$ (green dashed) tracks $P_{\mathrm{req}}$ closely for the vast majority of time steps, indicating that the proactive load reduction strategy successfully avoids curtailment in most periods. The six shaded intervals mark peak demand periods. The two inset panels zoom into the curtailment events on February 23 and February 29. On February 23 (left inset, 14:30-16:30), the TSO curtails approximately 50–100 MW over a two-hour window. On February 29 (right inset, 14:00-20:00), curtailment persists over a longer duration but with smaller magnitude. Outside these events, $P_{acc}$ and $P_{req}$ are virtually indistinguishable.

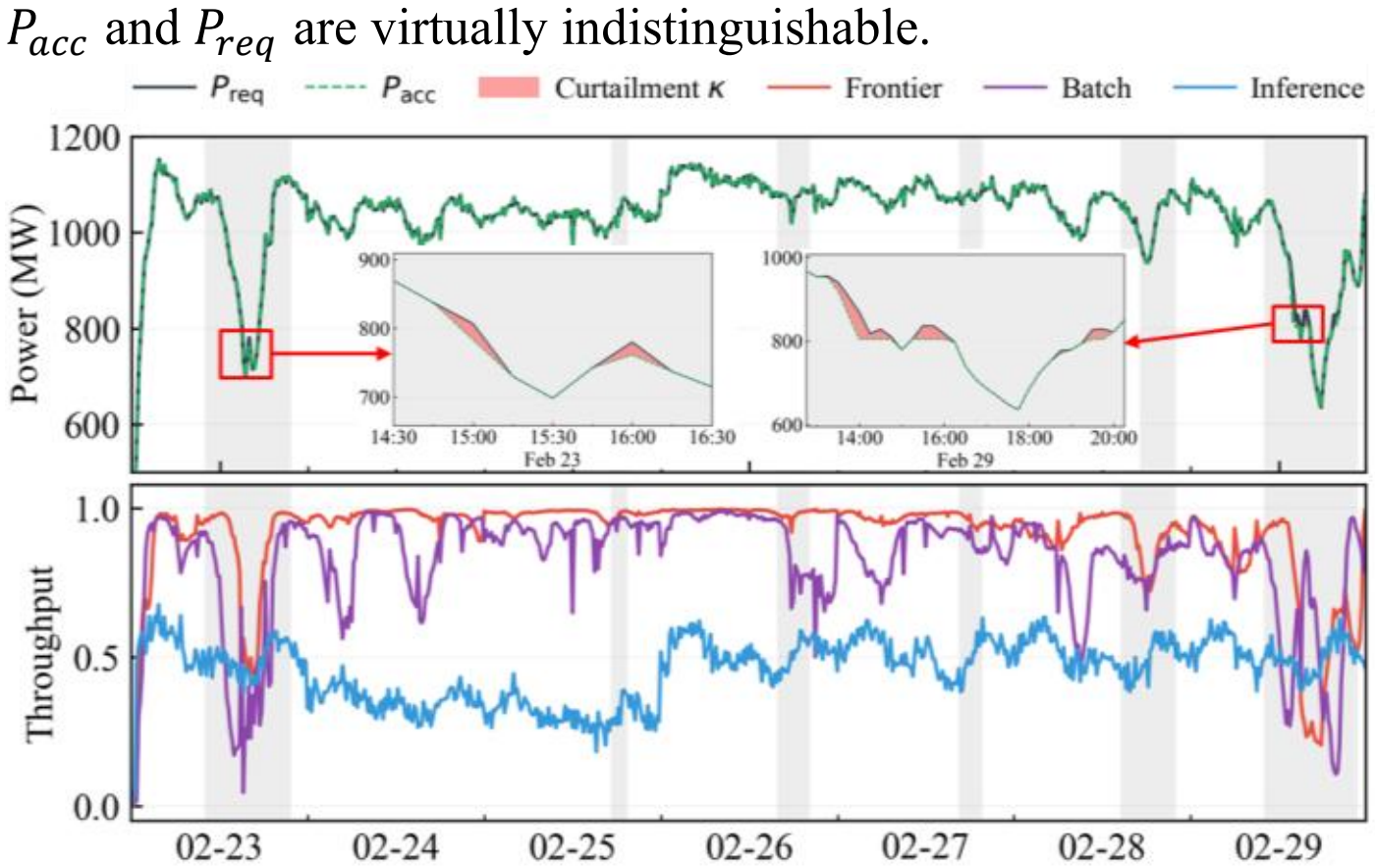

Fig. 3. Weekly operation of the SAC policy between February 23, 2024 and February 29, 2024. (Upper) power request, accepted exchange, and curtailment events. (Lower) cluster-level throughput allocation.

The lower panel reveals how the three clusters collectively contribute to the grid-responsive power modulation. Frontier throughput (red) operates between 0.8 and 1.0 under normal conditions and maintains relatively high throughput even when system demand exceeds the 75th percentile, with deep reductions occurring only on February 23 and 29, reaching a minimum of approximately 0.3. Batch throughput (purple) exhibits a more pronounced response: it fluctuates between 0.5 and 1.0 over the week and drops to 0.05-0.2 during the peak demand periods. Compared to Frontier, which sustains high throughput throughout the week, the SAC policy preferentially curtails Batch output when grid stress arises, resulting in visibly larger fluctuations in the purple trace. Inference throughput (blue) follows an independent diurnal cycle with peaks near 0.5-0.6 and troughs near 0.1-0.2, showing no visible correlation with the peak demand shading. To quantify this differentiated response, the evaluation period is partitioned into peak periods (system demand above the 75th percentile) and off-peak periods (below the 25th percentile). Table III summarizes the cluster-level behavior, where Δ denotes the off-peak value minus the peak value.

TABLE III: SAC CLUSTER BEHAVIOR DURING PEAK AND OFF-PEAK PERIODS

| Metric | Peak | Off-peak | Δ |
|---|---|---|---|
| Mean $P_{\mathrm{req}}$ (MW) | 970 | 1,063 | +93 |
| Mean $\kappa$ (MW) | 2.18 | 0.00 | -2.18 |
| $s_{1a}$ (Frontier) | 0.81 | 0.97 | +0.16 |
| $s_{1b}$ (Batch) | 0.68 | 0.90 | +0.21 |
| $s_2$ (Inference) | 0.49 | 0.44 | -0.05 |

Batch training exhibits the largest throughput swing (Δ=0.21), confirming its role as the primary grid-elastic resource. Frontier training is also reduced during peaks (Δ=0.16) but more conservatively, maintaining 98% cumulative workload completion over the evaluation week. Inference throughput is independent of grid conditions (Δ=-0.05), tracking the exogenous demand rather than responding to system load. The differentiated scheduling across clusters is driven by the asymmetric penalty weights: reducing Batch incurs the lowest cost ($M_{1b}$), followed by Frontier ($M_{1a}$), while inference rejection carries the highest per-unit penalty ($\alpha_{\mathrm{rej}}$). The policy accordingly treats Batch as the primary absorption buffer, moderates Frontier only when necessary, and leaves Inference to track its exogenous demand.

### *D. Curtailment Response: Execution Reallocation and BESS Buffering*

Despite the proactive load reduction demonstrated in Section C, residual curtailment events remain unavoidable during the most severe grid stress periods. This section examines how the execution optimizer and the on-site BESS jointly manage these events, using the most stressed day of the evaluation period (February 29) as a detailed case. Fig. 4 shows the operation on February 29. The upper panel displays the stacked executed IT power alongside the planned cluster boundaries (dashed lines) and the total $P_{req}$ and $P_{acc}$. The lower panel shows the BESS charge/discharge power and SoC trajectory.

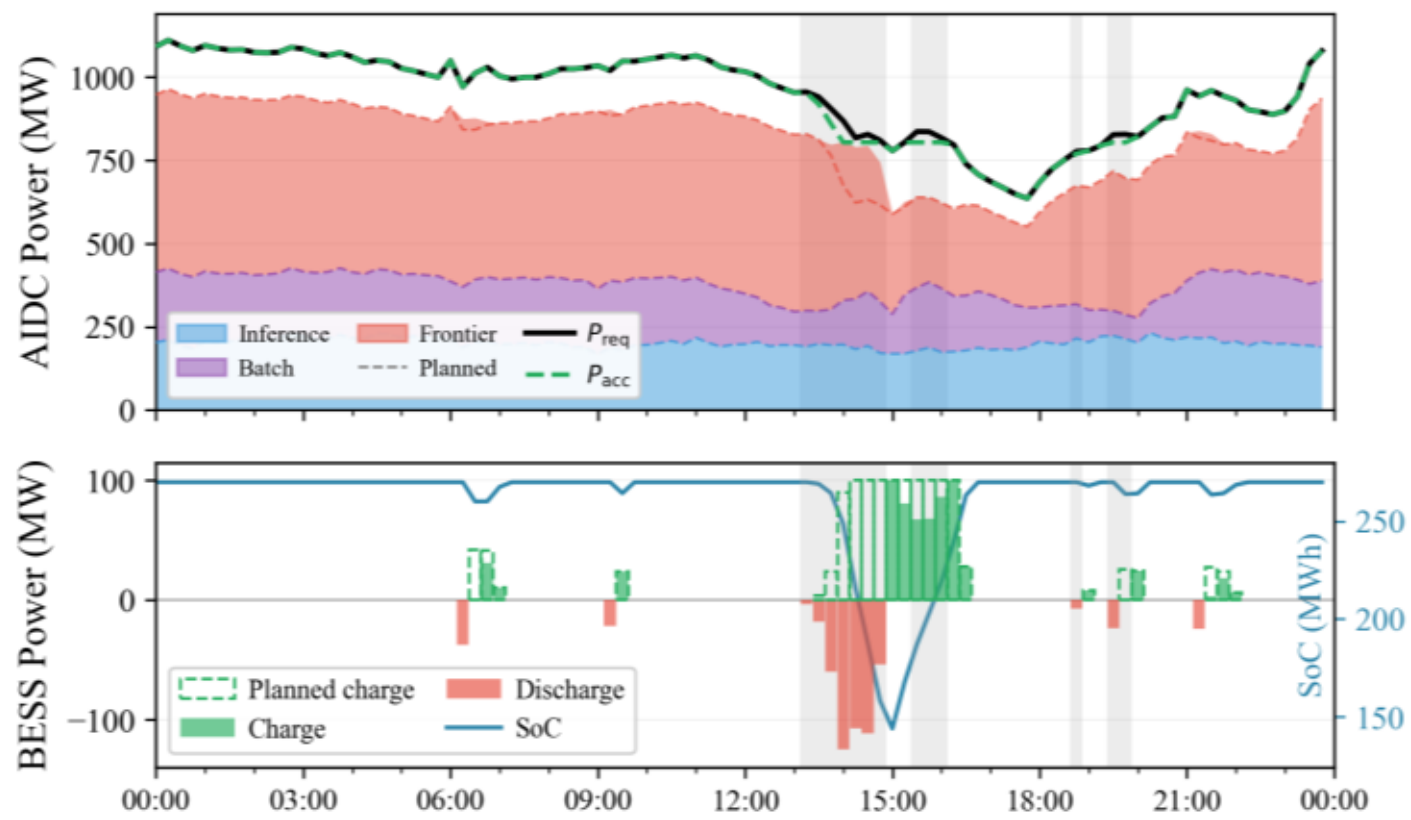

Fig.4 Curtailment response on February 29: cluster power reallocation (upper) and BESS dispatch (lower).

During the primary curtailment event (13:15-14:45, peak $\kappa$ = 64.8 MW at 14:00), the TSO caps $P_{acc}$ at approximately 804 MW. The planning layer and execution optimizer respond through two complementary mechanisms. First, the SAC policy proactively reduces its cluster throughput targets in the hours preceding the event, Batch throughput drops from 0.96 at midnight to 0.46 by 13:00, and Frontier from 0.98 to 0.97, lowering $P_{\mathrm{req}}$ from over 1,100 MW to 954 MW before curtailment begins. Second, once curtailment occurs, the BESS discharges progressively from 270 MWh to 143.8 MWh, delivering up to 125 MW to supplement the reduced $P_{\mathrm{acc}}$. The BESS discharge partially offsets the power shortfall, allowing the execution optimizer to maintain Frontier throughput above the level the planning layer had targeted. Of the 14 curtailment steps on this day, 9 are accompanied by BESS discharge.

Once the primary curtailment subsides at 15:00, the charging policy begins replenishing the SoC from 143.8 MWh back to 270 MWh. During the secondary curtailment events at 15:30-16:00 ($\kappa$ = 32.5, 32.0, 13.7 MW), the execution optimizer reduces the BESS charging power from the planned 100 MW to 67-86 MW, redirecting the freed capacity to sustain cluster throughput. The curtailment is absorbed entirely by the reduction in charging power, with no additional throughput loss beyond what the planning layer had already specified. The SoC fully recovers to 270 MWh by 16:45. This mechanism demonstrates a second mode of BESS buffering: in addition to active discharge during severe curtailment, the BESS provides passive support by deferring its own charging demand when the grid is constrained. Across the full evaluation week, the BESS remains idle for 92.6% of time steps, confirming its role as a curtailment spike buffer rather than a daily cycling resource.

### *E. TSO Curtailment Mechanisms and AIDC Workload Delivery*

The preceding sections evaluate the framework under a single TSO robustness configuration ($\Gamma_{\mathrm{U}}$=5, $\varepsilon$=0.07). In practice, the degree of demand uncertainty and network congestion varies across systems and operating conditions. This section examines how the framework performs as TSO-side stress parameters change, and how curtailment intensity propagates to AIDC workload delivery.

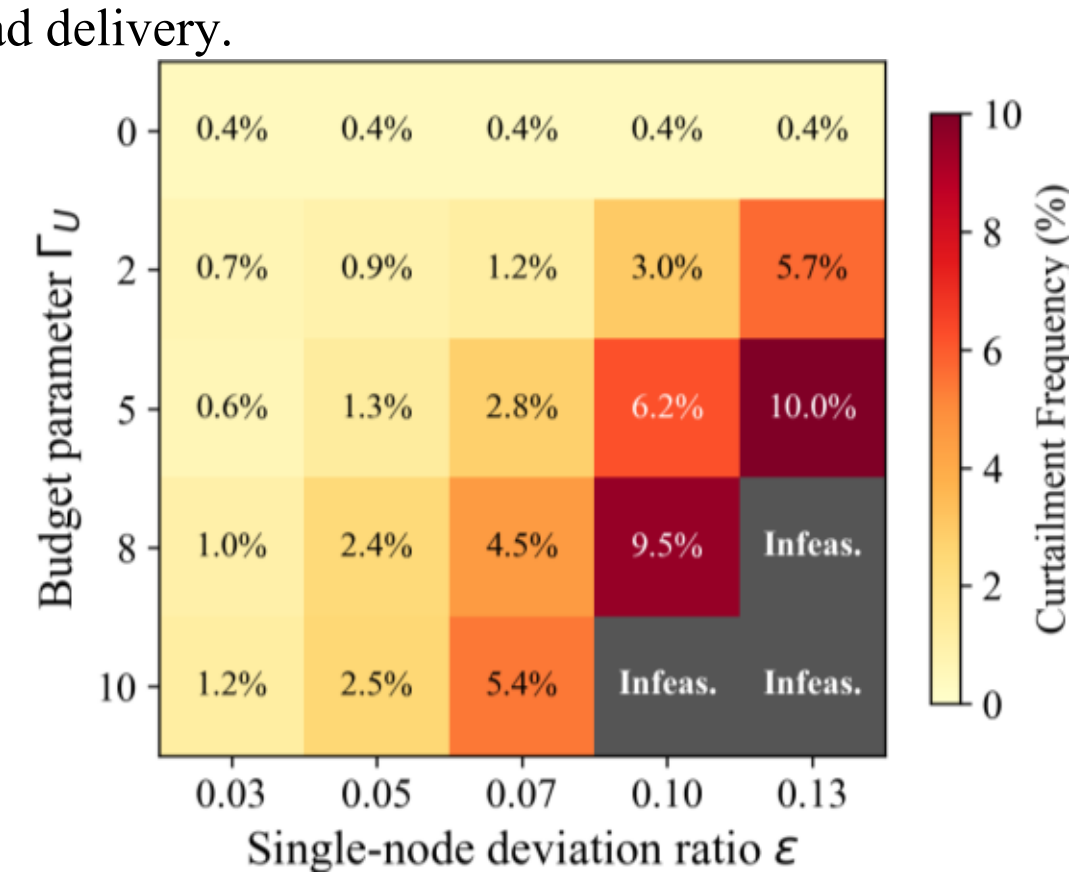

Fig.5 Curtailment frequency across budget parameter $\Gamma_{\mathrm{U}}$ and single-node deviation ratio $\varepsilon$.

Fig. 5 reports the curtailment frequency across 25 combinations of $\Gamma_{\mathrm{U}}$ and $\varepsilon$ . The $\Gamma_{\mathrm{U}} = 0$ row yields a uniform 0.4% curtailment across all values of $\varepsilon$, confirming that demand uncertainty alone does not trigger curtailment when the budget parameter is zero; the residual 0.4% is attributable to the PCC admission ramp limit ($R_{\mathrm{grid}}$=150 MW/step), which is physically unavoidable regardless of the AIDC's scheduling strategy. As $\Gamma_U$ increases, curtailment remains near this baseline for small $\varepsilon$ but rises sharply at larger $\varepsilon$, indicating that curtailment is driven by the joint effect of the budget parameter and the deviation ratio rather than by either parameter alone. At the default configuration ( $\Gamma_{\mathrm{U}} = 5$ , $\varepsilon = 0.07$ ), curtailment reaches 2.8%, and rises monotonically to 10.0% at ($\Gamma_{\mathrm{U}} = 5$, $\varepsilon = 0.13$). At the extreme corners ($\Gamma_{\mathrm{U}} \geq 8$, $\varepsilon \geq 0.13$ and $\Gamma_{\mathrm{U}} \geq 10$, $\varepsilon \geq 0.10$), the worst-case demand realization exceeds the grid's dispatch capability and the TSO acceptance problem becomes infeasible.

Fig. 6 traces the operational consequences of these curtailment levels on AIDC workload delivery, using five representative configurations spanning 0.6% to 10.0% curtailment frequency. The upper panel plots the cumulative curtailed energy over the evaluation week. Below 2.8%, cumulative curtailment remains modest and concentrated in a small number of isolated events. Above 5%, curtailment accumulates steadily throughout the week, with the 10.0% configuration reaching approximately 800 MWh by the end of the horizon. The lower panel shows the completion lag for Frontier (Cluster 1a) and Batch (Cluster 1b) training, defined as the percentage shortfall relative to a uniform delivery schedule. At 0.6% and 1.3% curtailment, both clusters track the uniform schedule closely, with completion lags remaining below 2% throughout. At the default 2.8%, a modest separation appears: Frontier lag stays within 3%, while Batch lag rises to approximately 5% by the end of the week, reflecting the planning layer's preferential protection of Frontier throughput. At 5.4% and 10.0%, the completion lags diverge sharply in the final days: Batch lag reaches approximately 10-15%, while Frontier lag remains comparatively contained, confirming that the emergent flexibility hierarchy persists under elevated grid stress.

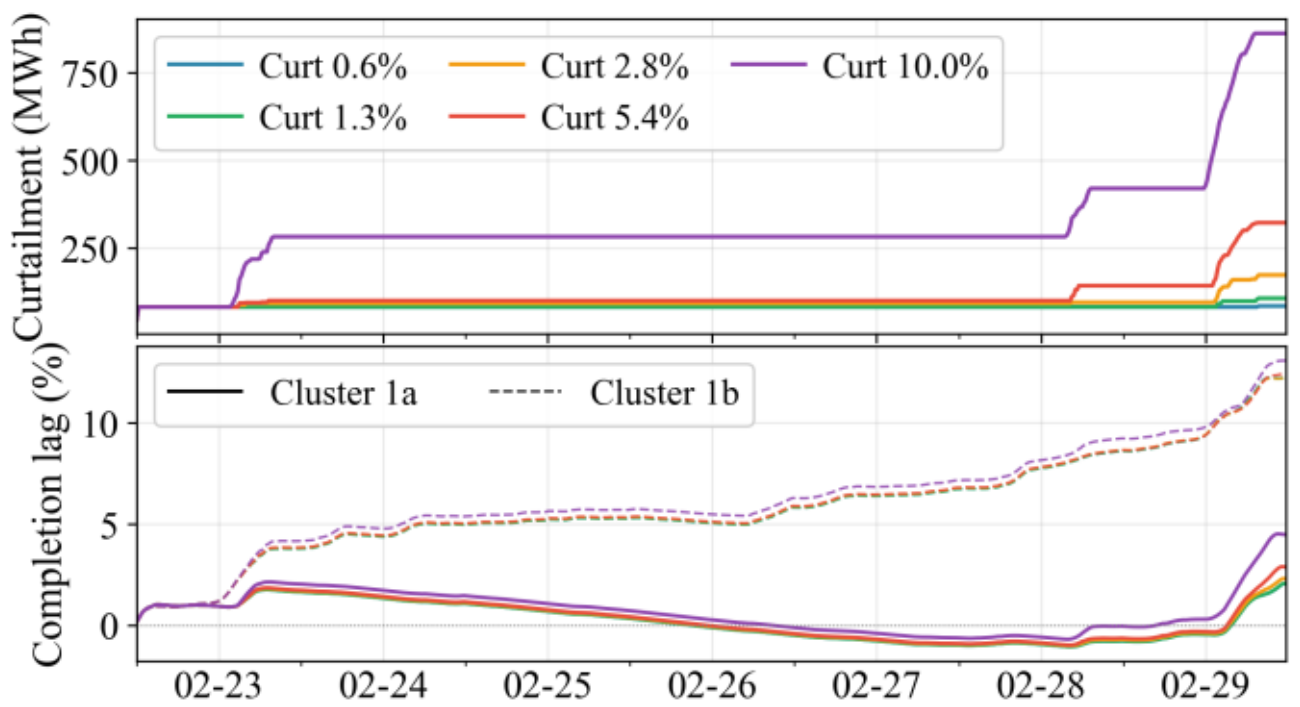

Fig.6 Cumulative curtailed energy (upper) and training workload completion lag (lower) under five curtailment intensity levels.

## VII. Conclusion

This paper develops an interaction framework for gigawatt-scale AIDCs operating under connect-and-manage. A sequential request–acceptance protocol formalizes the real-time AIDC–TSO coordination with an explicit curtailment variable and information boundary. Physical models on both sides of the PCC capture heterogeneous workload flexibility with on-site BESS and network feasibility under budget-constrained demand uncertainty. A three-layer hierarchical architecture organizes the closed-loop decision sequence under information asymmetry. Case studies demonstrate that the framework reduces curtailment from 9.1% to 2.8% while preserving 98.1% frontier training workload, that batch training acts as the primary grid-elastic resource while inference remains demand-driven, and that the on-site BESS provides curtailment buffering through active discharge and charge deferral.

Several limitations should be noted. The 15-minute scheduling resolution does not capture sub-minute power transients. The uncertainty set captures only demand-side deviations, whereas high renewable penetrations would introduce supply-side variability. Future work will address multi-AIDC coordination under AC power flow with N-1 security constraints, and improve policy generalization across grid topologies and market environments.